\documentstyle[epsfig,amssymb,mathptm]{mn}

\newcommand{\gta}{\ga}

\begin{document}
\def\lsim{\, \lower2truept\hbox{${< \atop\hbox{\raise4truept\hbox{$\sim$}}}$}\,}
\def\gsim{\, \lower2truept\hbox{${> \atop\hbox{\raise4truept\hbox{$\sim$}}}$}\,}

\title[Gravitational lensing]
      {Gravitational lensing: effects of cosmology and of 
lens and source profiles} 
\author[Perrotta F., Baccigalupi C., Bartelmann M., De Zotti G., Granato G.L.]
       {F. Perrotta$^{1,2}$,  C. Baccigalupi$^{2}$, M. Bartelmann$^{3}$,
        G. De Zotti$^{1}$, G. L. Granato$^{1}$ \\
        $^1$Oss. Astr. Padova, Vicolo dell'Osservatorio 5, I-35122
        Padova, Italy. Email: {\tt perrotta@sissa.it, dezotti@pd.astro.it,  
        granato@pd.astro.it}\\
        $^2$SISSA/ISAS, Astrophysics Sector, Via Beirut, 4, I-34014
        Trieste, Italy. Email: {\tt bacci@sissa.it}\\
        $^3$ Max-Planck Institut f\"ur Astrophysik, P.O.~Box 1317,
        D--85741 Garching, Germany. Email: {\tt
        msb@mpa-garching.mpg.de}}

\maketitle

\begin{abstract}

We present detailed calculations of the magnification 
distribution, including both weak and strong lensing, 
using very recent solutions of the Dyer-Roeder (1973) equation 
for light propagation in a inhomogeneous universe with a 
cosmological constant and    
up-to-date models for the evolving cosmological distribution of dark 
matter halos. We address in particular the effect on the magnification 
distribution  
of a non-zero cosmological constant, of different density profiles 
of lenses, and of finite sizes of lensed sources, 
three important issues not yet fully settled.

We show that, if dark matter fluctuations are normalized to
the local cluster abundance, in the presence of a cosmological constant
the optical depth for lensing {\em decreases\/} compared to the case of 
an Einstein-de Sitter universe, because
halos in the relevant mass range are less abundant over a large redshift
interval. We also discuss the differences in the magnification probability 
distributions produced by Navarro, Frenk \& White (NFW) and by Singular 
Isothermal Sphere (SIS) density profiles of lenses. We find that 
the NFW lens is more efficient for moderate magnifications ($2\lsim 
A\lsim 4$), and less efficient for larger magnifications. 
Moreover, we discuss quantitatively the maximum 
magnification, $A_{\rm max}$, that can be achieved in the case of extended 
sources (galaxies) with realistic luminosity profiles, taking into account 
the possible ellipticity of the lens potential. We find that plausible values 
of $A_{\rm max}$ are in the range 10--30. \\
Finally, we apply our results to a class of sources following the
luminosity evolution typical for a unified scheme of QSO formation. We
find that the lensed source counts at $850\,\mu$m can be larger than the 
unlensed ones by several orders of magnitude  at flux densities 
$\gta100\,$mJy. 
\end{abstract}

\section{Introduction}
\label{intro} 

Gravitational lensing is well known to be a powerful tool to 
probe the overall geometry of the universe at $z \lsim 6$ and 
the values of cosmological parameters (curvature, vacuum energy 
density described either by a cosmological constant or by a dynamical 
quantity like quintessence, Hubble constant: Bartelmann et al. 1997; 
Falco et al. 1998; Cooray 1999; Huterer \& Cooray 1999; Macias-Perez et al. 
2000; Bhatia 2000; Helbig 2000), the evolution of large scale structure 
(Rix et al. 1994; Mao \& Kochanek 1994; Bacon et al. 2000), the masses and the 
density profiles of dark halos of galaxies and galaxy clusters (Narayan 1998 
and references therein; Mellier 1999; Clowe et al. 2000).

Another important effect of gravitational lensing is the modification 
of the observed luminosity functions of distant sources and of their 
counts, due to the redshift dependent magnification bias (Peacock 1982; 
Narayan 1989; Schneider 1992).

In view of all the above applications,  it is essential to perform
the most accurate modeling of all the ingredients entering into the 
gravitational lensing process. Quantitative predictions depend on the lens 
population as a function of mass and redshift, on the lens profiles, 
on the the luminosity function and size of the sources, and on a correct 
treatment of the distance-redshift relations, taking into account 
the effect of density inhomogeneities on the propagation of light rays.  

Gravitational lensing by intervening mass concentrations is a
statistical process. It can be described by a probability distribution
$F(A,z_{\rm s})$ of the magnification factor $A$, which obviously depends
on the source redshift $z_{\rm s}$. It is generally difficult to model
$F(A,z_{\rm s})$ in detail analytically. Some authors estimated it
with $N$-body simulations (see e.g. Rauch 1991; Pei 1993a,b). 
Nonetheless, the main
features of $F(A,z_{\rm s})$ can be captured using generic properties of
gravitational lensing. Strong lensing occurs near caustic curves,
which cause $F(A,z_{\rm s})$ to decrease $\propto A^{-3}$ for $A\gg1$,
irrespective of the lens model (Peacock 1982; Turner et al. 1984). 

In this paper we present detailed calculations of the magnification 
distribution, including both weak and strong lensing, 
using very recent solutions of the Dyer-Roeder (1973) equation 
for light propagation in a inhomogeneous universe with a 
cosmological constant (Demianski et al. 2000) 
to compute the relationships between distances and redshift as well as 
up-to-date models for the evolving cosmological distribution of dark 
matter halos (Sheth \& Tormen 1999; Sheth et al. 1999) that can act as 
lenses. We address in particular the effect on the magnification  
distribution of a non-zero cosmological constant, of different density profiles 
of lenses, of a finite size and of luminosity profiles of lensed sources, 
three important issues not yet fully settled.
The magnification probability distribution has been applied to the  counts 
at $850 \mu m $: the luminosity function of the source population used 
here  has been inferred by the evolution of the QSOs luminosity function 
and from the link  between the QSOs and the hosting spheroids (Granato et 
al.~2000). Such luminosity function turns out to fit the SCUBA data.

The plan of this paper is as follows. In Sect.~\ref{lenses} and
\ref{densityprofile}, we describe our assumptions on the distribution
of lenses in mass and redshift, and on their density profiles, respectively. 
In Sect.~\ref{formalism}, we review the basic definitions of distance and
flux magnification in inhomogeneous universes. In
Sect.~\ref{distribution}, we introduce the probability for a source at
given redshift to be magnified by a certain amount, and quantify the 
magnification bias on source counts, for which a model is described in
Sect.~\ref{galaxy}. In Sect.~\ref{maxampli} we face the problem of
determining the maximum magnification allowed for extended sources. 
We summarize and discuss our results in Sect.~\ref{results}.

Throughout this paper, $\Omega_{\rm 0m}$ and $\Omega_{0\Lambda}$ denote the
present-day density parameters for the non-relativistic matter and for
the cosmological components, respectively, neglecting the radiation energy 
density. The Hubble constant is $H_0=100\,h\,{\rm km}\,{\rm s}^{-1}
\,{\rm Mpc}^{-1}$,
$\Omega_{0\cal R}\equiv1-\Omega_{\rm 0m}-\Omega_{0\Lambda} $ is the
curvature term, and $\Omega$ denotes the total density parameter. Our
fiducial model has $\Omega_{\rm 0m}=0.3$, $\Omega_{0\Lambda}=0.7$ and
$h=0.65$. Furthermore, $q_0=-\ddot{a}a/\dot{a}^2$ is the deceleration
parameter, and we concentrate on spatially-flat models ($\Omega_{0\cal
R}=0)$.

\section{The Evolving Halo Mass Function}
\label{lenses}

We assume that the lens population consists of collapsed dark matter halos 
with an epoch-dependent mass function described by the 
Sheth \& Tormen (1999) function, which reproduces fairly accurately  the 
results of extensive numerical simulations over more than four orders 
of magnitude in mass, for a wide range of Cold Dark Matter cosmologies 
(Jenkins et al. 2000). This function improves considerably on the familiar 
Press-Schechter (1974) model which overestimates the abundance of ``typical'' 
($M_\star$) halos and underestimates that of massive systems. 
The comoving number density of halos with mass $M$ at redshift $z$ is then
\begin{eqnarray}
  \frac{{\rm d}n}{{\rm d}M} &=& \sqrt{\frac{2aA^2}{\pi}}\,
  \frac{\rho_0}{M^2}\,\frac{\delta_{\rm c}(z)}{\sigma(M)}\,
  \left[1+\left(\frac{\sigma(M)}{\sqrt{a}\delta_{\rm c}(z)}
  \right)^{2p}\right]\nonumber\\ 
  &\times& \left|\frac{{\rm d}\ln\sigma}{{\rm d}\ln M}\right|\,
  \exp\left(-\frac{a\delta_{\rm c}^2(z)}{2\sigma(M)^2}\right)\;;
\label{Sheth}
\end{eqnarray}
The best-fit values of the parameters are $a=0.707$, $p=0.3$, and
$A\simeq 0.3222$ (Sheth \& Tormen 1999; Sheth, Mo \& Tormen 1999). 
The Press-Schechter mass function is recovered for
$a=1$, $p=0$ and $A=0.5$.  

In Eq.~(\ref{Sheth}), $\rho_0$ is the mean mass density at a
reference epoch $t_0$, which we assume to be the present time, and $\sigma^2$ 
is the variance of linear density fluctuations at the present epoch 
smoothed with a spherical top-hat filter $W_R(k)$ enclosing a mass $M$. 
The variance $\sigma^2$ is related to the power spectrum $P(k)$ by:
\begin{equation}
\label{variance}
  \sigma^2(R)={4 \pi \over (2 \pi) ^3} \, \int_0^\infty{\rm d}k\, 
 k^2\,W_R^2(k)\,P(k)\;, 
\end{equation}
with
\begin{equation}
\label{filter}
  W_R(k)=\frac{3}{(kR)^3}\,[\sin(kR)-(kR)\cos(kR)]\;.
\end{equation} 
%
%
A useful expression for $\sigma(M)$ and its derivative ${\rm
d}\sigma/{\rm d}M$ was obtained by Kitayama \&
Suto (1996),
\begin{equation}
  \sigma\propto(1+2.208m^d-0.7668m^{2d}+0.7949m^{3d})^{-2/(9d)}\;,
\end{equation}
$\Gamma=\Omega_{0m}  h \exp[-\Omega_{0b}(1+\sqrt{2h} /
\Omega_{0m} ) ]$ being the shape parameter (Bardeen et al. 1986) and  
we have adopted a present baryon density $\Omega_{0b} h^2 =0.03$.
For a $\Lambda$CDM model with
$\Omega_{0\Lambda}=0.7$ and $\Omega_{\rm 0m}=0.3$, the COBE
normalization is $\sigma_8=\sigma(R=8\,h^{-1}\,{\rm Mpc})=0.925$
(Bunn \& White 1997). 

Also in Eq.~(\ref{Sheth}), $\delta_{\rm c}^2(z)$ is the linear density
contrast of an object virializing at $z$, evaluated at the present
epoch. It can be estimated using the spherical collapse model
(e.g.~Peebles 1980; Lahav et al.~1991; Lacey \& Cole 1993; Nakamura
1996; Eke, Cole \& Frenk 1996; {\L}okas \& Hoffman 2000) yielding: 
\begin{equation}
\label{deltac}
  \delta_{\rm c}(z)=\frac{\delta_{\rm c}\,D_+(z=0)}{D_+(z)}\;,
\end{equation}
where $D_+(z)$ is the linear growth factor of density fluctuations:
\begin{eqnarray}
  D_+(z) &\simeq& (1+z)^{-1} {5 \over 2}\,\Omega_{\rm m}\nonumber\\ 
  &\times& [\Omega_{\rm m}^{4/7}-\Omega_\Lambda+
  (1+\Omega_{\rm m}/2)(1+\Omega_\Lambda/70)]^{-1}\;,
\end{eqnarray}
in the approximation by Carroll et al.~(1992), where $\Omega_{\rm
m}=\Omega_{\rm 0m}(1+z)^3/E^2(z)$, $\Omega_\Lambda=\Omega_{0\Lambda}
/E^2(z)$, and
\begin{equation}
\label{ez}
  E(z)=H(z)/H_0=[\Omega_{\rm 0m}(1+z)^3+\Omega_{0\cal R}(1+z)^2+
  \Omega_{0\Lambda}]^{1/2}\;.
\end{equation}
$\delta_{\rm c}$ [Eq.~(\ref{deltac})] is the value of the critical
overdensity evaluated at virialization. An approximate
expression valid for any cosmological model was reported by Nakamura
(1996) and Kitayama \& Suto (1996):
\begin{equation}
  \delta_{\rm c}\approx\frac{3\,(12\pi)^{2/3}}{20}\,
  (1+0.0123\log_{10}\Omega_{\rm m})\;.
\end{equation}

\section{Gravitational Lensing and Halo Models}
\label{densityprofile}

The ray-tracing equation relates the position of a source to the
impact parameter in the lens plane of a light ray connecting source
and observer. The light ray passing the lens at an impact parameter
$\vec\zeta$ is bent by an angle $\hat{\alpha}(\vec\zeta)$. The source
position $\vec\eta$ and the impact parameter $\vec\zeta$ in the lens
plane are related through
\begin{equation}
  \vec\eta=\frac{D_{\rm s}}{D_{\rm d}}\,\vec\zeta-
  D_{\rm ds}\,\vec{\hat{\alpha}}(\vec\zeta)\;,
\end{equation} 
where $D_{\rm d,s,ds}$ are the angular-diameter distances between
observer and lens, observer and source, and lens and source,
respectively. We will specify these distances later in the context of
a ``clumpy universe''. The deflection angle $\vec{\hat{\alpha}}$ is
the sum of the deflections due to all mass elements of the lens
projected on the lens plane,
\begin{equation}
\label{hatalpha0}
  \vec{\hat{\alpha}}(\vec\zeta)=\frac{4G}{c^2}\,\int_{R^2}\,
  \frac{\vec\zeta-\vec\zeta'}{\left|\vec\zeta-\vec\zeta'\right|^2}\,
  \Sigma(\vec\zeta')\,{\rm d}^2\zeta'\;,
\end{equation} 
where $\Sigma$ is the surface mass density of the lens.

It is convenient to introduce the dimensionless impact parameter 
$\vec x=\vec\zeta/\zeta_0$, where $\zeta_0$ is an arbitrary length scale. 
The corresponding value projected
in the source plane is $\eta_0=\zeta_0\,D_{\rm s}/D_{\rm d}$ and the source 
position is $\vec y=\vec\eta/\eta_0$.

The dimensionless surface mass density $\kappa(x)$, also called {\em
convergence\/}, is
\begin{equation}
  \kappa(x)=\frac{\Sigma(x)}{\Sigma_{\rm cr}}
  \quad\mbox{with}\quad
  \Sigma_{\rm cr}=\frac{c^2}{4\pi G}\,
  \frac{D_{\rm s}}{D_{\rm d}D_{\rm ds}}\;.
\end{equation}
The dimensionless ray-tracing equation is then,
\begin{equation}
\label{lensequation0}
  \vec y=\vec x-\vec\alpha(\vec x)\;,
\end{equation} 
with the scaled deflection angle
\begin{equation}
\label{scaledalpha}
  \vec\alpha(\vec x)=\frac{D_{\rm d}D_{\rm ds}}{\zeta_0\,D_{\rm s}}\,
  \vec{\hat{\alpha}}(\zeta_0\,\vec x)=
  \frac{2}{x}\,\int\,{\rm d}x'\,x'\,\kappa(x')\;.
\end{equation}
The axial symmetry of the lenses considered here allows us to rewrite the
lens equation in terms of the dimensionless ``mass'' $m(x)$ enclosed
within the radius $x$,
\begin{equation}
\label{lensequation}
  \vec y=\vec x-\frac{m(x)}{x}\;,
\end{equation} 
with
\begin{equation} 
  m(x)=2\,\int_0^x\,{\rm d}x'\,x'\,\kappa(x')\;.
\end{equation} 
The Jacobian matrix ${\cal A}(\vec x)=\partial\vec y/\partial\vec x$
for the lens mapping [Eq.~(\ref{lensequation})] determines the magnification
factor for each image $i$,
\begin{equation}
  \mu_i=\frac{1}{\det{\cal A}(\vec x_i)}\;.
\end{equation}
The sign of $\mu_i$ reflects the parity of the image with respect to
the source. The total source magnification is the sum of the absolute
values of the magnification factors of all images,
\begin{equation} 
\label{totalampli}
  \mu=\sum_i|\mu_i|\;.
\end{equation} 
For axially symmetric lenses, the Jacobian determinant of the lens
mapping can also be written
\begin{eqnarray}
\label{zeri}
  \det{\cal A}&=&\left(1-\frac{\alpha(x)}{x}\right)\,
  \left(1-\frac{{\rm d}}{{\rm d}x}\alpha(x)\right)\nonumber\\
  &=&\left(1-\frac{m}{x^2}\right)\,
  \left[1-\frac{{\rm d}}{{\rm d}x}\,\left(\frac{m}{x}\right)\right]
  =(1-\kappa)^2-\gamma^2\;,
\end{eqnarray}
where $\gamma$ is the shear (cf.~Schneider, Ehlers \& Falco ~1992). Critical
curves are located at the zeros of $\det{\cal A}$; their images in
the source plane under the mapping described by Eq.~(\ref{lensequation0}) 
are the caustics.

A simple model for the mass profile of a lens (cluster or galaxy) is
the singular isothermal sphere (SIS; e.g.~Binney \& Tremaine 1987):
\begin{equation}
  \rho(r)=\frac{\sigma_v^2}{2\pi Gr^2}\;,
\label{rhoSIS}
\end{equation}   
where $\sigma_v$ is the line-of-sight velocity dispersion. 
In this case, the deflection angle is independent of the impact parameter
(cf.~Schneider et al. 1992; Narayan \& Bartelmann 1997).
One of the two critical curves degenerates to a point. Any given
source has either one or two images. Two images appear only if the
source lies inside the Einstein ring. The Einstein angle corresponding
to a SIS is
\begin{equation}
\label{thetae}
  \theta_{\rm E}=\hat{\alpha}\,\frac{D_{\rm ds}}{D_{\rm s}}
\label{thetaE}
\end{equation}
with 
\begin{equation}
\label{hatalphaSIS}
  \hat{\alpha}=4\pi\frac{\sigma_v^2}{c^2}=1.4''\,
  \left(\frac{\sigma_v}{220\,{\rm km\,s}^{-1}}\right)^2\;.
\end{equation}
In order to evaluate the mass of a halo with velocity dispersion 
$\sigma_v$ from Eq.~(\ref{rhoSIS}), we truncate the sphere at an 
effective radius computed following  Lahav et al. (1991).

If every halo virializes to form a singular isothermal sphere, mass
conservation implies that the velocity dispersion is related to the
mass and to the virialization redshift by (see, e.g.~Porciani \& Madau 2000; 
Kaiser 1986): 
\begin{equation}
\label{velocitydispersion0}
  \sigma_v=\frac{1}{2}\,H_0\,r_0\,\Omega_{\rm 0m}^{1/3}\Delta^{1/6}\,
  \left[\frac{\Omega_{\rm 0m}}{\Omega_{\rm m}}\right]^{1/6}\,
  (1+z)^{1/2}
\end{equation}
where $\Omega_{\rm m}$ is defined in the previous section, 
$r_0=(3M/4\pi\rho_0)^{1/3}$,  and $\Delta(z)$ is the mean density of
the virialized halo in units of the critical density at
that redshift.  A useful expression for the dependence of $\sigma_v$
on redshift is given in Bryan \& Norman (1998),
\begin{equation}
\label{velocitydispersion}
  \sigma_v=M^{1/3}\,[H^2(z)\Delta(z)G^2/16]^{1/6}\;.
\end{equation}
Bryan \& Norman (1998) also give a fitting formula for $\Delta(z)$ for
a flat universe with a cosmological constant and for a universe with
$\Omega_{0\cal R}\ne0$ and $\Omega_\Lambda=0$. For a flat universe,
($\Omega_{0\cal R}=0$) their hydrodynamical simulations yield: 
\begin{equation}
\label{deltalambda}
  \Delta(z)=18\pi^2+82x-39x^2, 
\end{equation}
where $x=\Omega_{\rm m}-1$.

The SIS model is useful because it allows to work out analytically 
the basic lensing properties. On the other hand, 
high-resolution $N$-body simulations (Navarro, Frenk
\& White 1997) showed that in hierarchically clustering
universes, virialized dark matter halos have a universal density
profile (referred to as NFW), which is shallower than isothermal near the 
center and steeper in the outer regions: 
\begin{equation}
\label{NFW}
  \rho(x)=\frac{\rho_{\rm crit}\delta_{\rm NFW}}{x\,(1+x)^2}\;,
\end{equation}
where $x=r/r_{\rm s}$, $r_{\rm s}$ being a scale radius, $\delta_{\rm
NFW}$ is the characteristic density contrast of the halo, and
$\rho_{\rm crit}$ is the critical density at the epoch of the
halo virialization. This formula was found to accurately describe the 
equilibrium density profiles of dark matter halos over a broad range of masses
($3\times10^{11}\le M_{200}/M_\odot\le3\times10^{15}$), irrespective of
the cosmological parameters and the initial density fluctuation
spectrum.

Still higher resolution $N$-body simulations indicate a steeper central cusp 
than that of the NFW profile: $\rho(x)\propto 
\left[x^{1.5}(1+x)^{1.5}\right]^{-1}$ (Moore et al. 1999; Ghigna et al. 2000). 
Since the slope of the central density profile in this case [$\rho(r)\propto 
r^{-1.5}$] is intermediate between those of the NFW [$\rho(r)\propto 
r^{-1}$] and of the SIS [$\rho(r)\propto r^{-2}$], the two cases 
considered here (NFW and SIS) will bracket it.  

We parameterize halos by their mass $M_{200}$ (defined as the 
mass within the virial radius $r_{200}$, the radius of a sphere of 
mean interior density $200\rho_{\rm crit}$; see Navarro et al. 1997), 
since this is the mass whose distribution is given by 
Eq.~(\ref{Sheth}). Unless otherwise specified, we abbreviate
$M_{200}$ by $M$. The halo concentration is $c=r_{200}/r_{\rm s}$. It
is related to the density parameter $\delta_{\rm NFW}$ by
\begin{equation}
\label{deltavsc}
  \delta_{\rm NFW}=\frac{200}{3}\frac{c^3}{[\ln(1+c)-c/(1+c)]}\;.
\end{equation} 
The virial radius of a halo at redshift $z$ depends on the halo mass as
\begin{equation}
\label{virialradius}
  r_{200}=\frac{1.63\times10^{-2}}{(1+z)}\,
  \left(\frac{M}{h^{-1}\,M_\odot}\right)^{1/3}\,
  \left[\frac{\Omega_{\rm 0m}}{\Omega_{\rm m}(z)}\right]^{-1/3}\,
  h^{-1}\,{\rm kpc}\;.
\end{equation}
Alternatively, the halo can be characterized by its circular velocity,
\begin{eqnarray}
\label{velocityvirial}
  V_{200}&=&\left(\frac{GM_{200}}{r_{200}}\right)^{1/2}\nonumber\\
  &=&
  \left(\frac{r_{200}}{h^{-1}\,{\rm kpc}}\right)\,
  \left[\frac{\Omega_{\rm 0m}}{\Omega_{\rm m}(z)}\right]^{1/2}\,
  (1+z)^{3/2}\,{\rm km\,s}^{-1}\;.
\end{eqnarray}
The scale radius $r_{\rm s}$ depends on the halo mass. The halo
concentration increases with decreasing halo mass (e.g.~Fig.~6 of
Navarro et al. 1997). Less massive halos are therefore more
concentrated.  For a given halo mass, Eqs.~(\ref{virialradius}) and
(\ref{deltavsc}) completely specify the density profile [Eq.~(\ref{NFW})].

The lens equations for the NFW profile are given by Bartelmann (1996)
and Maoz et al.~(1997). The surface mass-density is
\begin{equation}
\label{surfacedensity}
  \Sigma(x)=\frac{2\rho_{\rm crit}\delta_{\rm NFW}\,r_{\rm s}}
  {x^2-1}\,f(x)\;,
\end{equation}
with
\begin{equation}
  f(x)=\left\{\begin{array}{ll}  
    1-{2 \over \sqrt{ x^2-1} } \arctan\sqrt{ (x-1) \over (x+1)} &
    \mbox{($x>1$)} \\
    1-{2 \over \sqrt{1-x^2}} {\rm arctanh}\sqrt{(1-x) \over (x+1)}  &
    \mbox{($x<1$)} \\
    0 & \mbox{(x=1)}
\end{array}\right.\;.
\end{equation}
The dimensionless lens mass is
\begin{equation} 
  m(x)=\frac{4\rho_{\rm crit}\delta_{\rm NFW}\,r_{\rm s}}
        {\Sigma_{\rm cr}}\,g(x)\;,
\end{equation} 
with
\begin{equation} 
\label{gx}
  g(x)=\ln\frac{x}{2}+1-f(x)\ .
\end{equation} 
While the NFW profile [Eq.~(\ref{NFW})] has two critical curves (Bartelmann
1996), the SIS profile has only one. A SIS lens has either one or two
images, an NFW lens either one or three.

\section{The Magnification Bias}
\label{formalism}

In this section, we follow the approach by Schneider et al. 
(1992) and Schneider (1987a,b). We call $\mu$ the magnification of
fluxes in an inhomogeneous universe, and relate it to the
magnification $A$ in a homogeneous universe.

The effect of lensing on flux-limited source counts is quantified by
the magnification bias (Turner et al. 1984). If the surface density
of galaxies with flux greater than $S_\nu$ is
$N(S_\nu)$, flux magnification alters the counts to
$N'(S_\nu)=\mu^{-1}\,N(S_\nu\mu^{-1})$, and the
magnification bias is given by (Narayan 1989):
\begin{equation}
\label{bias}
  q(\mu,S_\nu)=\frac{N'}{N}=
  \frac{N(S_\nu\mu^{-1})}{\mu\,N(S_\nu)}\;.
\end{equation}
The factor $\mu^{-1}$ arises because solid angles are magnified, hence
source counts are diluted. Eq.~(\ref{bias}) implies that 
for power-law counts ($N\propto S_\nu^{-\alpha}$), $q(\mu,S_\nu)
=\mu^{\alpha-1}$, so that the 
magnification bias increases as the counts steepen.

Since the reference to ``unlensed'' counts has given rise to confusion
in the literature, it is important to specify our model for light
propagation in the universe when talking about the number of sources
seen ``in absence of lensing''.

In Eq.~(\ref{bias}), $N$ is defined as the number count 
of sources that do not appear behind lenses. Because of
energy conservation, the flux of these objects must be reduced compared
to a homogeneous universe with the same average density. In
other words, since lensing is caused by inhomogeneities,
we have to deal with an inhomogeneous universe. While sources observed
through lenses are magnified by a factor $\mu^+>1$ compared to a
homogeneous universe, unlensed sources must
be demagnified by $\mu^-<1$. 
When comparing lensed to unlensed sources, the effective magnification
is
\begin{equation}
\label{effective}
  \mu=\frac{\mu^+}{\mu^-}>1
\end{equation}
(e.g.~Schneider 1984, 1987a,b). This is the ``empty beam approach'' to
light propagation (Dyer \& Roeder 1973; Ehlers \& Schneider 1986), in
which light cones are devoid of clumped matter. The fraction of uniformly 
distributed matter is denoted by $\alpha_{\rm s}$, the smoothness parameter.

In this {\em on-average\/} homogeneous and isotropic universe, the
average flux $\langle S\rangle$ from a source population at redshift
$z$ with luminosity $L$, must be equal to the corresponding flux
$S_{\rm FL}$ that would be observed in a strictly uniform 
Friedman-Lema{\^\i}tre universe,
\begin{equation}
\label{meanflux}
  \langle S\rangle=S_{\rm FL}=
  \frac{L\,K(L,z)}{4\pi\bar{D}_{\rm L}^2(z)}\;,
\end{equation}
where $\bar{D}_{\rm L}(z)$ is the luminosity distance in the uniform  
universe, and $K(L,z)$ is the K-correction. In general, it is not
possible to compare mean fluxes of sources in these two different
spacetimes. In particular, the notion of distance has no unique
meaning in a clumpy universe, as it depends on both redshift and
direction. Due to the corrugated structure of the gravitational field,
the propagation of light in the inhomogeneous universe is a
statistical problem. We can compare fluxes as in Eq.~(\ref{meanflux})
only because we {\em assume\/} that the global geometry of the clumpy
universe equals that of the homogeneous
universe. Equation~(\ref{meanflux}) implies that the area of a surface
of constant redshift on the future light cone of a source is equal to
that of the corresponding wavefront in the Friedman-Lema{\^\i}tre 
model. In order to use redshift-distance relations in an inhomogeneous, 
partly clumpy universe, which is uniform on average, it is then necessary to 
introduce a meaningful operational definition of 
distance. Since light propagation in a clumpy universe depends on
the clumps in and near the light beam, Dyer \& Roeder (1973)
considered the limiting case of a light bundle that propagates far
away from all clumps and is thus unaffected by gravitational
shear. This is the ``empty cone'' limit.

The angular-diameter distance is then computed by replacing the
density parameter $\Omega_{\rm m}$ with the ``homogeneous'' fraction
$\alpha_{\rm s}\Omega_{\rm m}$. The resulting ``Dyer-Roeder'' 
distance is larger than the angular-diameter distance in a homogeneous
universe. For a spatially-flat universe with deceleration parameter
$q_0=0.5$, the Dyer-Roeder distance between redshifts $z_1$ and $z_2>z_1$ is
\begin{equation}
\label{DR}
  r(z_1,z_2,\beta)=\frac{2}{\beta}\,\left[
    \frac{(1+z_2)^{(\beta-5)/4}}{(1+z_1)^{(\beta+5)/4}}-
    \frac{(1+z_1)^{(\beta-5)/4}}{(1+z_2)^{(\beta+5)/4}}
  \right]\;,
\end{equation}
where $\beta=(1+24\Omega_{\rm G})^{1/2}$ and $\Omega_{\rm
G}=(1-\alpha_{\rm s})\Omega_{\rm m}$ is the density parameter in
compact objects ($\beta=\alpha_{\rm s}=1$ for the homogeneous
universe).  Hereafter, we abbreviate $r(0,z,\beta)\equiv r(z,\beta)$
and $r(z,1)\equiv r_1(z)$. The dimensional angular diameter distance
in a homogeneous universe is
\begin{equation}
  D_1(z)=\frac{c}{H_0}\,r_1(z)\;.
\label{D_A}
\end{equation}
The problem is more complicated when the cosmological constant is
positive.  The Dyer-Roeder equation in a FRW universe with non-zero
cosmological constant has been solved exactly by Kantowski (1998),
Kantowski \& Kao (2000) and, following a different approach, by
Demianski et al.~(2000).
%
%
The general solution can be expressed in terms of the hypergeometric
functions $f_{s\pm}$ introduced by Demianski et al.~(2000):
\begin{equation}
  r(z,\beta) = A_1\,\frac{(1+z)^{-\beta/4}}{(1+z)^{5/4}}\,
  \frac{f_{s+}}{(1+z)^3}+
  A_2\,\frac{(1+z)^{\beta/4}}{(1+z)^{5/4}}\,
  \frac{f_{s-}}{(1+z)^3}\;,
\label{DRlambda}
\end{equation} 
where the constants $A_{1,2}$ are determined by the boundary
conditions
\begin{equation}
  \left.r(z,\beta)\right|_{z=0}=0\quad\mbox{and}\quad
  \left.\frac{{\rm d}r}{{\rm d}z}\right|_{z=0}=1\;.
\end{equation}
This can be used to generalize the two-point distance of 
Eq.~(\ref{DR}), which, for a non-zero cosmological constant, reads:
\begin{eqnarray}
  r(z_1,z_2,\beta) &=& r(z_1,\beta)(1+z_1)r(z_2,\beta)\nonumber\\
  &\times&\int_{z_1}^{z_2}{\rm d}z'\,
  \frac{[(1+z')^3\Omega_{\rm 0m}+\Omega_{0\Lambda}]^{1/2}}
       {r^2(z',\beta)(1+z')^2}\;,
\label{DRlambdaz1z2}
\end{eqnarray}  
with $r(z,\beta)$ given by Eq.~(\ref{DRlambda}).

For practical applications, these exact solutions are rather
cumbersome to use. Demianski et al.~(2000) give approximate solutions
for $r(z,\beta)$ and $r(z_1,z_2,\beta)$, which we use in this paper in
the form kindly provided by R.~de Ritis (private communication) for
$\Omega_{0\Lambda}=0.7$ and $\alpha_{\rm s}=0.9$, implying
$\beta=1.84$, and $\Omega_G=0.1$. We ignore a possible time-dependence
of the smoothness parameter. The effect of different choices of
$\alpha_{\rm s}$ on the Dyer-Roeder distance with positive
cosmological constant is shown in Demianski et al.~(2000).

Having introduced the Dyer-Roeder distance, we can interpret the magnification
in Eq.~(\ref{effective}) as the ratio between the flux $S$ actually
received from a source at redshift $z$, and the flux $S_{\rm empty}$
that would be received if the same source was observed through an
empty cone:
\begin{equation}
\label{mu}
  \mu=\frac{S}{S_{\rm empty}}=
  S\,\frac{4\pi D_{\rm L}^2(z,\beta)}{L\,K(L,z)}>1\;,
\end{equation}
where 
\begin{equation}
\label{lumdist}
  D_{\rm L}(z,\beta)=\frac{c}{H_0}\,(1+z)^2\,r(z,\beta)
\end{equation}
is the luminosity distance in an empty cone.

Since the clumpy universe is homogeneous on-average,
$\bar{D}_{\rm L}(z,\beta)=D_{\rm L}(z,1)$. Using Eq.~(\ref{mu}),
Eq.~(\ref{meanflux}) reads
\begin{equation}
\label{mumean}
  \langle\mu\rangle=\left(\frac{r(z,\beta)}{r_1(z)}\right)^2>1\;.
\end{equation}
$\langle\mu\rangle$ can be interpreted as the magnification of an
average light beam in a smooth universe relative to that for an empty
cone in a clumpy universe: $\langle\mu\rangle=S_{\rm FL}/S_{\rm
empty}$.

For $q_0=1/2$, $\langle\mu\rangle$ becomes
\begin{equation}
  \langle\mu\rangle={\beta}^{-2}\,
  \frac{{\rm sinh}^2[0.25\beta\ln(1+z)]}
       {{\rm sinh}^2[0.25\ln(1+z)]}\;,
\end{equation}
(Pei 1993b), while for the more general case including the cosmological
constant, $\langle\mu\rangle$ has to be derived from
Eq.~(\ref{mumean}).

Let $p(\mu,z){\rm d}\mu$ denote the probability for a source at
redshift $z$ to be magnified by a factor of $\mu$ within ${\rm d}\mu$. 
The normalization and flux conservation conditions require:
\begin{equation}
\label{condmu}
  \int_1^\infty\,{\rm d}\mu\,p(\mu,z)=1\;,\quad
  \int_1^\infty\,{\rm d}\mu\,p(\mu,z)\mu=\langle\mu\rangle\;.
\end{equation}
When dealing with source counts on cosmological scales, however, it is
often convenient to refer magnifications to the homogeneous universe
rather than to the ``demagnified'' background.  We call $A$ the
magnification relative to the homogeneous universe: 
\begin{equation}
  A=\frac{S}{S_{\rm FL}}=\frac{S_{\rm empty}}{S_{\rm FL}}\,\mu\;,
\end{equation} 
where $\mu=S/S_{\rm empty}$ is the magnification of a source seen
through an empty beam, as in Eq.~(\ref{mu}).
Eqs.~(\ref{meanflux}) and (\ref{mumean}) imply: 
\begin{equation}
\label{adimu}
  A=\frac{\mu}{\langle\mu\rangle}\;.
\end{equation}
This is the definition of magnification used, for example, by
Blain (1996) and Peacock (1982). It is trivial to show that for
a given source redshift $z$, $p(\mu,z){\rm d}\mu=p(A,z){\rm d}A$, and
the normalization and flux conservation conditions become
\begin{equation}
\label{condA}
  \int_{A_{\rm min}}^\infty{\rm d}A\,\, p(A,z)=1\;,\quad
  \int_{A_{\rm min}}^\infty{\rm d}A\,\, p(A,z)A=1\;.
\end{equation}
Here, $A_{\rm min}$ is the minimum magnification relative to the FRW
universe, that can be produced by compact lenses, i.e.~$A_{\rm
min}=\langle\mu\rangle^{-1}$. Since $\langle A\rangle=1$, the sources
which are demagnified with respect to the homogeneous universe 
have $A<1$. 

\section{The Magnification Distribution}
\label{distribution}

The number of images produced by a lens, their angular separation and
their magnification depend on their relative alignment as seen by the
observer. For a fixed geometry of the lens system, one can ask where
the source must lie for its images to have a total magnification
larger than a given value $\mu$. The area of the resulting region in
the source plane is the cross section $\sigma(\mu,z_{\rm d},z_{\rm
s},\chi)$, which obviously depends on lens and source redshifts,
$z_{\rm d}$ and $z_{\rm s}$, on a set of parameters $\chi$ describing
the lens model, and on the magnification itself. As both SIS and NFW 
halos are completely characterized by
their mass, $\chi=M$ here. The cross section quantifies the efficiency
of the individual lens on a source. It is generally found numerically,
by solving the lens equation and finding the area in the source plane
where a source must lie in order to produce images with a total
magnification larger than $\mu_{\rm tot}$.  The cross section of a
singular isothermal sphere for a total magnification $\mu_{\rm
tot}>\mu$ is
\begin{equation}
\label{sigmaSIS} 
\sigma(\mu_{\rm tot}>\mu)=\frac{4\pi{\hat{\alpha}}^2\,D_{\rm
ds}^2}{\mu^2}  \quad\mbox{for}\quad\mu\ge2\;.
\end{equation}  
We need to compute the total magnification cross section of an
ensemble of lenses distributed according to the mass function of 
Eq.~(\ref{Sheth}). We denote by $n_{\rm c}(z,M)$ the comoving lens number
density, and $n(z,M)\equiv(1+z)^3 n_{\rm c}(z,M)$. As long as 
the cross sections of individual lenses do not overlap, each 
light bundle from a source encounters only one lens, and the total
cross section is the integral of the individual cross sections over
the redshift and mass distributions of lenses:
\begin{eqnarray}
  \sigma_{\rm tot}(\mu,z_{\rm s})&=&4\pi\left(\frac{c}{H_0}\right)^3
  \\ &\times&
  \int_0^{z_{\rm s}}\,{\rm d}z\int{\rm d}M\,
  \frac{\sigma(\mu,z,z_{\rm s},M)\,n_{\rm c}(z,M)\,
        (1+z)^2\,r_1^2(z)}
       {\sqrt{\Omega_{\rm 0m}(1+z)^3+\Omega_{0\Lambda}}}\;,\nonumber
\end{eqnarray}
where the proper volume of a spherical shell of width ${\rm d}z$ at
redshift $z$ is
\begin{equation}
  {\rm d}V=4\pi\left(\frac{c\,r_1(z)}{H_0}\right)^2
  \times\frac{{\rm d}r_{\rm prop}}{{\rm  d}z}\,{\rm d}z\;.
\end{equation}
For a spatially-flat  universe
\begin{equation}
  \frac{{\rm d}r_{\rm prop}}{{\rm d}z}=\frac{c}{H_0\,(1+z)\,
  \sqrt{\Omega_{\rm 0m}(1+z)^3+\Omega_{0\Lambda}}}\;.
\end{equation}
The probability for a source at redshift $z_{\rm s}$ to be lensed with
magnification $>\mu$ is obtained by dividing $\sigma_{\rm tot}$ by the
area of the source sphere
\begin{eqnarray}
  P(\mu,z_{\rm s}) &=& \frac{c}{H_0}\,\frac{1}{r_1^2(z_{\rm s})}\,
  \int_0^{z_{\rm s}}\,{\rm d}z\,
  \frac{r_1^2(z)\,(1+z)^2}
       {\sqrt{\Omega_{\rm 0m}(1+z)^3+\Omega_{0\Lambda}}}\nonumber\\ 
  &\cdot&
  \int{\rm d}M\,\sigma(\mu,z,z_{\rm s},M)n_{\rm c}(z,M)\;.
\label{Prob}
\end{eqnarray}
The requirement of non-overlapping cross sections restricts the
validity of Eq.~(\ref{Prob}) to $P\ll1$, i.e.~the total cross
section must be much smaller than the area of the source sphere.  The
net effect of gravitational lensing on the distribution of flux
densities expected from a population of distant sources can be
described by the probability distribution of magnifications,
$p(\mu,z)$ introduced in Sect.~\ref{formalism}. While $P\ll1$, the
differential probability is $p(\mu,z)=-{\rm d}P(\mu,z)/{\rm d}\mu$; in
particular, $p(A,z)=p(\mu,z)\langle\mu\rangle$.

The differential probability decreases as $\mu^{-3}$ for $\mu\gg1$,
hence the high magnification tail in terms of $A$ can be written as
$p(A,z)\propto a(z)A^{-3}$. On the other hand, Eq.~(\ref{Prob}) breaks
down for small magnifications, where multiple lensing events become
important and cross sections overlap (in fact, the probability
for many low-magnification lensing events along the line of sight to a
source is rather large, while a single interaction producing high
magnifications is a relatively rare event). In particular, this
implies that there is a critical magnification $A_{\rm cut}$ below
which multiple lensing becomes important, resulting in
low-magnification events (weak lensing regime). 
Vietri \& Ostriker (1983) analytically described multiple
lensing for the SIS model.

Based on general considerations (see Bartelmann \&
Schneider 2001 for a review), one expects a skewed 
magnification distribution with a weak lensing peak near $A=1$, turning 
into the high-magnification tail $\propto A^{-3}$ at $A=A_{\rm cut}$.
 
For a Gaussian density fluctuation field $\delta$, weak lensing by
large scale structure (e.g.~Bartelmann \& Schneider 2001;
Miralda-Escud\'e 1991; Blandford et al.~1991; Kaiser 1992; Jain \&
Seljak 1992) produces a Gaussian magnification distribution.  In fact,
as long as $\delta A\equiv A-1\ll1$, the magnification of a source at
redshift $z$ can be approximated as
\begin{equation}
  A(z)=1+\delta A(z)\approx1+2\kappa(z)\;,
\end{equation}
i.e.~to first order the magnification
fluctuation is just twice the convergence $\kappa$, which itself is a
projection of the density contrast $\delta$. The distribution of the
magnification fluctuations $\delta A$ will then be Gaussian, with mean
zero and a dispersion $\sigma_{\rm A}(z)$ which depends on the source
redshift and, albeit weakly, on cosmology. Typical values for
$\sigma_{\rm A}$ run from $\sim2\times10^{-3}$ for $z=0.05$ to
$\sim0.44$ at $z=7.5$ (cf.~Bartelmann \& Schneider 2001).

A convenient choice for $A_{\rm cut}$ is $A_{\rm cut}=1+1.5\sigma_{\rm A}(z)$, 
yielding $A_{\rm cut}\approx1.5-2$ for the redshift range of interest in this
paper. We model the probability distribution for $A<A_{\rm cut}$
as
\begin{equation}
\label{weakpdiA}
  p(A,z)=H(z)\,\exp[-(A-\bar{A})^2/2\sigma_{\rm A}^2(z)]\;,
\end{equation}
where the precise location of the peak $\bar{A}$ and the amplitude
$H(z)$ are determined by the normalization and flux conservation
conditions [Eq.~(\ref{condA})] on the {\em combined\/} (weak plus strong 
lensing) probability distribution.

The magnification probability for isothermal galaxy models has been
derived by Peacock (1982) and Vietri \& Ostriker (1983), and, for more
complicated galaxy models, by Blandford \& Kochanek (1987) and
Wallington \& Narayan (1993). In this paper, we integrate
Eq.~(\ref{Prob}) also for lenses with NFW density profile (see
Sect.~\ref{densityprofile}), and we describe the low-magnification
distribution according to Eq.~(\ref{weakpdiA}). Numerical results are
presented in Section~\ref{results}.

Let us now turn to the magnification bias on a flux-limited source
sample. The integrated counts above a flux density threshold $S_\nu$ of sources 
with a comoving luminosity function $\Phi(L,z)$ can be written as (see 
e.g. De Zotti et al.~1996):
\begin{equation}
\label{sourcecounts}
  N(S_\nu)=\int_0^{z_0}\,{\rm d}z\int_{L_{\rm min}}^\infty\,{\rm d}L\,
  \Phi(L,z)\,\frac{D_{\rm L}^2(z,1)}{(1+z)^2}\,
  \frac{{\rm d}r}{{\rm d}z} \qquad \hbox{sr}^{-1} ,
\end{equation}
where the luminosity distance $D_{\rm L}^2(z,1)$ is computed from 
Eq.~(\ref{lumdist}) with $\beta=1$ and 
\begin{equation}
\label{Lmin}
  L_{\rm min}(\nu)=4\pi(1+z)\,r^2(z)\,S_\nu\,
  \frac{L(\nu)}{L[(1+z)\nu]}\;.
\end{equation}
Note that the comoving radial coordinate $r$ [and its element $dr$ in 
Eq.~(\ref{sourcecounts})] must not be confused with 
the dimensionless angular diameter distance $r_1(z)$ 
[Eq.~(\ref{D_A}))].

The luminosity function modified by the magnification bias reads 
(e.g.~Pei 1995):
\begin{equation}
\label{phieffective}
  \Phi'(L,z)=\int_{A_{\rm min}}^\infty{\rm d}A\,
  \frac{p(A,z)}{A}\,\Phi\left(\frac{L}{A},z\right)\;.
\end{equation}
Allowance for the effect of lensing on counts is made by 
replacing $\Phi'(L,z)$ to $\Phi(L,z)$ in Eq.~(\ref{sourcecounts}).

\section{Source Counts in the Submillimetre Waveband}
\label{galaxy}

In this paper, the ``unlensed'' galaxy counts in the submillimetre
waveband are taken from the model presented by Granato et al.~(2000),
which is in good agreement with the available SCUBA data at $850\,\mu$m
(Blain et al.~1999; Smail et al.~1999). 

In this model the rate of formation of spheroids at high redshift is
estimated exploiting the (i) QSO Luminosity Function and (ii) observational
evidence leading to the conclusion that high redshift QSOs did shine in
the core of early type proto--galaxies during their main episode of star
formation. In their scenario the star formation is more rapid in more
massive objects, ranging from $\sim 0.5$  to $\sim 2$ Gyr when going from
more massive to less massive objects. This {\it Anti-hierarchical Baryonic
Collapse} is expected to occur in Dark Matter halos, when the processes
of cooling and heating is considered. The larger the dark halo and the
enclosed spheroid mass are, the shorter will the gas infall and cooling
times be, resulting in a faster formation of the stars and of the central
black hole. The star-formation process and the quasar shining phase
proceed until powerful galactic winds are caused by the quasar itself,
which occurs at a characteristic time when its luminosity becomes high
enough. Also Monaco et al (2000), in order to account for the observed
statistics of QSOs and elliptical galaxies in the framework of hierarchical
structure formation, introduced a a time delay decreasing with mass
between the beginning of the star formation and the QSO bright phase.

The spectroscopic evolution of galaxies adopted here is based on the model
GRASIL (Silva et al.\ 1998). GRASIL includes: ({\it i}) chemical
evolution; ({\it ii}) dust formation, assumed to follow the chemistry of
the gas; ({\it iii}) integrated spectra of simple stellar populations
(SSP) with the appropriate chemical composition; ({\it iv}) realistic 3D
distribution of stars, molecular clouds (in which stars form and
subsequently escape) and diffuse dust; ({\it v}) radiative transfer
computation in this clumpy ISM and dust temperature distribution
determined by the local radiation field.

With these ingredients, the evolving luminosity functions (LF) at
various wavelengths in the millimeter and submillimetre wavebands are
evaluated numerically, and they turn out to be significantly different
from models of Pure Luminosity Evolution  for the $60\,\mu$m LF of
IRAS galaxies (Saunders et al.~1990), properly rescaled to the
wavelengths of interest. The IRAS galaxy LF, which is based on an
empirical model describing the evolution in a parametric way, was used
by Blain (1996) to obtain galaxy counts in the submillimetre
wavebands, which were then used for estimating the incidence of
gravitational lensing on source counts. However, the model by Granato
et al.~implies steeper source counts, nearly exponentially decreasing
at bright fluxes, so that, as explained in section (\ref{formalism}),
the effect of gravitational lensing is expected to be more important.

All source counts obtained in this paper, with and without magnification
bias, the spheroids include elliptical galaxies as well as bulges of Sa
galaxies, and we followed the formalism of Granato et al.~(2000) on the
source properties.

\begin{figure}
\psfig{file=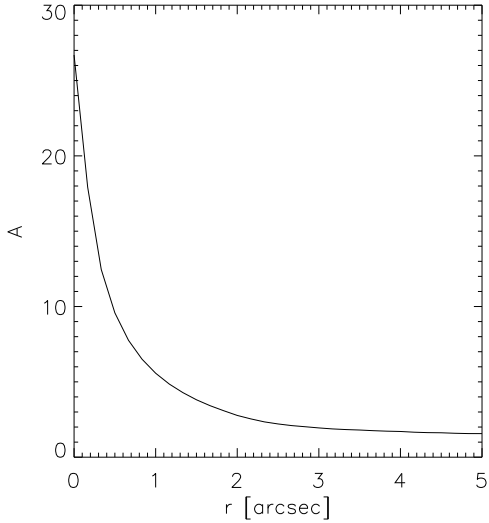,height=3.0in}
\caption{Magnification of an extended source as a function
of the offset \protect\( r\protect \) between its center and the
projection of the lens center in the source plane. Here the lens potential
has no ellipticity. See text for details. }
\label{fig:ag0}
\end{figure}

\begin{figure} 
\psfig{file=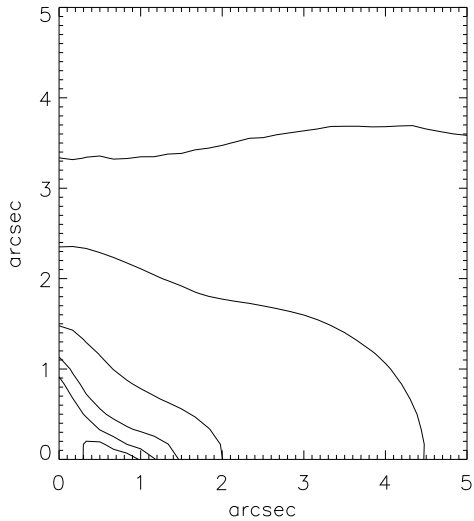,height=3.0in}
\caption{Contour plot of the magnification for an extended source when
\protect\( g=0.1\protect \). The projection of the lens center in the source
plane is at the origin of axes. The levels are at \protect\( A=1.5,2,4,6,8,10\protect \)
from outside in. See text for details.}
\label{fig:ag01}
\end{figure}

\section{Maximum magnification for extended sources}
\label{maxampli}
As discussed by Peacock (1982), in the case of extended sources the 
magnification cannot be arbitrarily large. Correspondingly the 
magnification 
distribution must be cut-off at large $A$ as: 
\begin{equation}
\label{correction}
  p(A)\propto \exp\left(-\frac{A}{A_{\rm max}}\right)\;, 
\end{equation}
where the cut-off magnification $A_{\rm max}$ depends on the physical size 
of the source. 

Since, in some applications, e.g. for the estimate of the influence of 
lensing on counts of sub-millimeter sources (Blain 1996), the results are 
sensitive to the adopted value of $A_{\rm max}$, the approximated 
expression derived by Peacock (1982) 
[$A_{\rm max}=70(DH_0/c)(d/\hbox{kpc})^{-1}$, 
where $d$ is the physical radius of the source, $D$ is the angular diameter 
distance of the source and $c/H_0$ is the Hubble radius] may not be 
sufficient.

The morphology of strongly lensed sources indicates that most lenses are
not circularly symmetric (e.g. Narayan \& Bartelmann 1997 and references
therein). Therefore, to estimate the maximum possible magnification for a
source of physical radius \( r \) at redshift \( z_{s} \), we considered
in general an elliptical lensing potential due to a quasi isothermal
sphere (Blandford \& Kochanek 1987). Defining polar coordinates  
\( r \) and \( \theta  \) in the
image plane, with origin at the center of mass, 
and measuring \( \theta  \) from the major axis of the ellipse, the 
deflection potential may be written as:

\begin{equation}
\label{eq:pot}
\Psi (r,\theta )= \theta_{\rm E} \sqrt{(s^{2}+r^{2})}(1-g\cos 2\theta )\ .
\end{equation}
Here \( g \) is the ellipticity parameter, \( s \) the core radius and 
$\theta_{\rm E}$ is defined by Eq.~(\ref{thetaE}).
The results
for an extended source are weakly dependent on \( s \), within a relatively 
broad interval;  therefore we simply set \( s=0 \) in the following. 
With this potential
we have computed the expected magnification of a pointlike source \(
A(\vec{y}) \), as a function of its position \( \vec{y} \) in the source
plane. This has been done by means of the ray-shooting method (e.g.\
Schneider et al. 1992, pag. 304). Then the magnification of an
extended source with brightness profile \( I(\vec{y}) \), as a function of
the position \( \vec{y}_{E} \) of its center, is given by
\begin{equation}
A_{E}(\vec{y}_{E})=\frac{\int I(\vec{y})A(\vec{y})\, 
d^{2}y}{\int I(\vec{y})\, d^{2}y} \ .
\end{equation}
For the brightness profile, we have used either a de Vaucouleurs law
\begin{equation}
\label{eq:dv}
\log I(R)=\log I_{e}-3.33\left[ (R/R_{e})^{1/4}-1\right] \ ,
\end{equation}
or the Hubble profile
\begin{equation}
I(R)=I_{o}/(1+R/R_{o})^{2} \ .
\end{equation}
Observed profiles of spheroidal galaxies are well reproduced by both
functional forms over a wide range of \( R \), provided that \( R_{e}\simeq
11\, R_{o} \) (Mihalas \& Binney 1981). 

We find that, adopting this relationship between the scale-lengths, the
estimated \( A_{E} \) are very similar in both cases. Therefore in the
following we present only results for Eq.~(\ref{eq:dv}). We assume in
particular \(R_{e}=5 \) kpc, typical for a bright elliptical galaxy (e.g.
Kent 1985). For a SIS, the maximum possible magnification is achieved
when the lens is local, i.e. when \( D_{\rm ds}=D_{\rm s} \) 
in Eq.~(\ref{thetaE}). We set
\( \sigma_v =300 \) km s\( ^{-1} \), which corresponds to a mass of \(
10^{12} \) M\( _{\odot } \) within 25 kpc adopting the isothermal
relationship \( M(<R)=2\sigma_v^2 R/G \). With our adopted cosmology, 
Eq.~(\ref{thetaE}) yields \( \theta_E=2.6'' \). 

Figure \ref{fig:ag0} shows the corresponding magnification for an extended
source as a function of the offset \( r \) between its center and the
projection of the lens center in the source plane. Here we have set \( g=0
\) into Eq.~(\ref{eq:pot}), i.e. we adopt a spherical potential, and
the source is at \( z_{s}=3 \), but very similar plots are obtained for \(
z_{s} \) in the range \( 1\div 4 \). As can be seen, the magnification is
maximized when the source and the lens are aligned and \( A_{\rm max}\simeq
26 \). If the lens is placed instead at \( z_{l}=0.5 \), then \( b=1.2 \)
arcsec and \( A_{\rm max}\simeq 13 \).

However, as already remarked, strongly lensed sources are usually
explained in terms of potentials with non vanishing ellipticity. A typical
value for \( g \) could be \( 0.1 \). In this case the symmetry of \(
A_{E} \) around \( r=0 \) is lost and the maximum 
magnification, \( A_{\rm max}\simeq 12 \), significantly lower than in the 
spherical case, occurs when
the lens and the source are offset by \( \sim 0.7'' \) along the
major axis of the ellipse. The results are detailed in Fig. \ref{fig:ag01}.

We conclude that reasonable values of $A_{\rm max}$ for extended sources 
are in the range 10--30. The lower value is a relatively conservative lower 
limit, easily exceeded for a wide range of values for the relevant 
quantities, while the latter is obtained only under rather special conditions.

\begin{figure}
\psfig{file=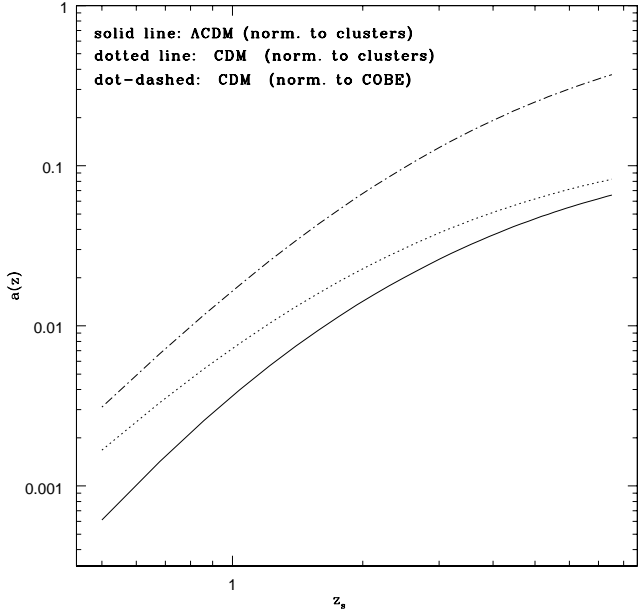,height=3.5in}
\caption{SIS model: comparison of the high-magnification tail
amplitudes $a(z)$ for a flat $\Lambda$CDM model with
$\Omega_{\Lambda}=0.7$ (solid line), and a flat CDM model (dotted
line), both normalized to the cluster abundance.  The amplitude is
plotted vs.~source redshift. The dot-dashed line refers to a
COBE-normalized flat CDM universe.}
\label{azpaper}
\end{figure}

\begin{figure}
\psfig{file=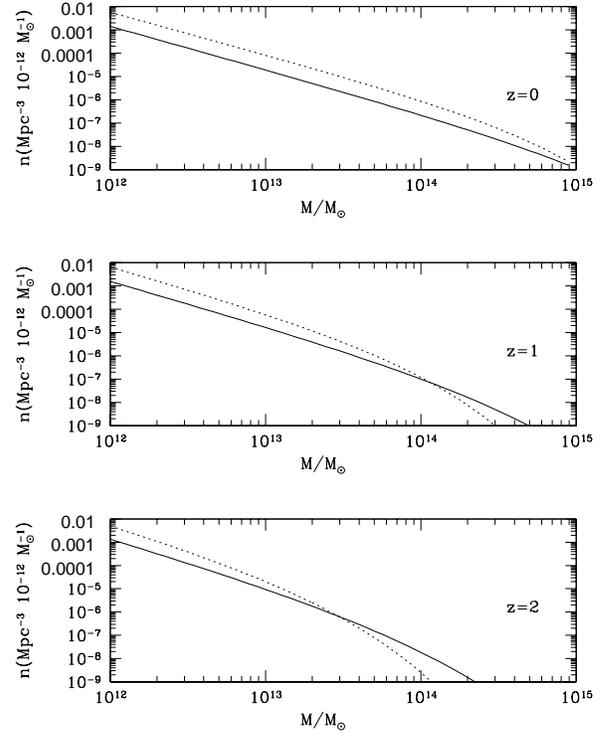,height=4.in}
\caption{The Sheth \& Tormen (1999) mass function for a flat CDM model
(dotted line) and for a $\Lambda$CDM model (solid line), at redshifts
$z=0,1,2$, respectively, from top to bottom.}
\label{nM012} 
\end{figure}

\begin{figure}
\psfig{file=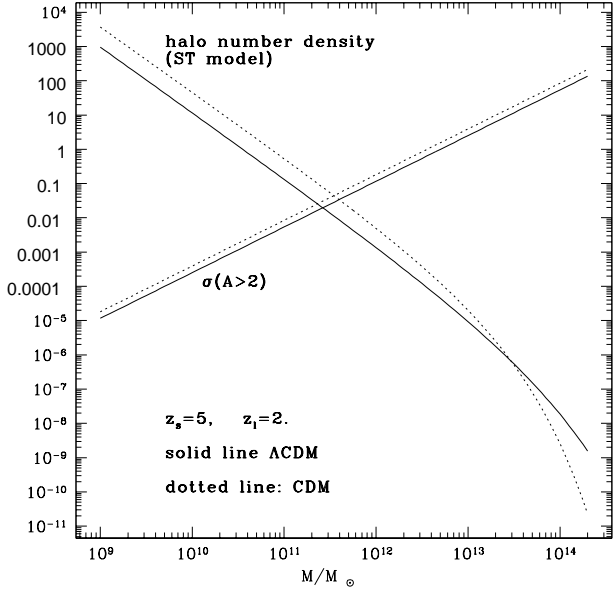,height=4.in}
\caption{Number density of dark-matter halos (from the Sheth \&
Tormen mass function), and cross section for magnification $A>2$ by a
SIS lens, in arbitrary units, as a function of the lens mass in
$M_\odot$. Dotted lines correspond to flat CDM, solid lines to
$\Lambda$CDM, both normalized to the cluster abundance.}
\label{factors} 
\end{figure}

\begin{figure}
\psfig{file=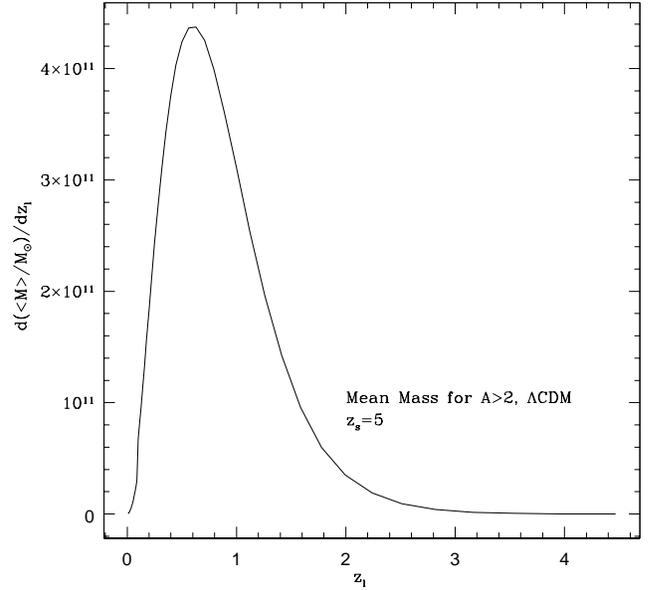,height=4.in}
\caption{Contributions from different redshifts to the effective lens mass 
(see text) for a fixed configuration of the lens
system, SIS lenses, and a $\Lambda$CDM model.}
\label{mmedia} 
\end{figure}

\section{Numerical Results and Discussion}
\label{results}

\subsection{Effects from cosmology}
\label{effect}



To illustrate the effect of the cosmological model on the magnification 
probability distribution $p(A,z)$, we plot in Fig.~\ref{azpaper} the 
amplitude $a(z)$ of the high-magnification tail for two different
cosmologies, viz.~a flat CDM and a $\Lambda$CDM model with
$\Omega_{0\Lambda}=0.7$ and $\Omega_{\rm 0m}=0.3$. We have adopted SIS profiles 
for lenses and a smoothness parameter $\alpha_{\rm s}=0.9$. The two models 
are normalized to reproduce the local abundance of rich clusters: 
$\sigma_8=0.56\Omega_{\rm 0m}^{-0.47}$ (Viana \& Liddle 1999). 

The optical depth for a beam of light from a source due to lensing is 
proportional to the number density of deflectors multiplied by the 
cross section for a given magnification, integrated along the 
line of sight. Two competing effects are therefore relevant. On one hand, 
the path length to a source is larger for a $\Lambda$CDM model. On the 
other hand, the structure formation histories within the standard hierarchical 
clustering scenario are also different for the different cosmologies. 
If we normalize the models to reproduce the local cluster abundance, the 
density of lower mass objects is lower in the $\Lambda$CDM  model, 
because the mass function is flatter 
(e.g.~Eke et al. 1996). This is shown in the upper panel of 
Fig.~\ref{nM012}, where the mass function [Eq.~(\ref{Sheth})] is plotted 
as a function of the halo mass at $z=0$. The number of low-mass objects keeps
increasing with redshift above the $\Lambda$CDM model, while the opposite is 
true for high masses. 

The next question is then: which is the mass range contributing most to 
the optical depth? In Fig.~\ref{factors} we show as a function 
of the lens mass, both the cross section for magnifications $A>2$ and 
the lens mass function for given values of the source and lens redshifts. 
A SIS profile is adopted. It may be noted that the magnification cross 
section is similar for the two models (although slightly higher for the 
flat CDM model). 
The product of the two functions peaks at masses between $10^{11}$ and 
$10^{12}\,\hbox{M}_\odot$ for both a flat CDM and a $\Lambda$CDM model. 

The effective mass of dark matter halos contributing most to strong 
lensing of a source located at $z_s$ can be estimated as:
\begin{equation}
\label{meanmass}
  \langle M\rangle=\int{\rm d}z_{\rm l}\int {\rm d}M\,M\,
  \frac{{\rm d}P(A,z_s)}{{\rm d}z_{\rm l}{\rm d}M}\;,
\end{equation}
where $P(A,z_s)$ is the probability to have magnification $> A$. The 
inner integral, ${\rm d}\langle M\rangle/{\rm d}z_{\rm l}$, is plotted 
as a function of $z_{\rm l}$, for $z_s=5$, a SIS lens profile and a 
$\Lambda$CDM model, in Fig.~\ref{mmedia}. As illustrated by this Figure, 
the maximum contribution to the magnification probability comes from the 
mass range ($10^{11}$--$10^{12}\,M_\odot$) 
for which space densities implied by a CDM model are appreciably higher 
than in the case of a $\Lambda$CDM model, in the relevant redshift interval, 
if the models have to be consistent with the observed cluster abundance. 
This over-compensates for the larger path length to a source in a $\Lambda$CDM 
model and explains why the probability distribution of strong 
magnifications  has a {\it lower} amplitude for such model.

It may be noted (Fig.~\ref{azpaper}) that the difference between the two 
cases decreases with increasing source redshift, since also the $\Lambda$CDM 
model approaches an Einstein-de Sitter universe at high redshift.

Also shown in Fig.~\ref{azpaper} is the amplitude $a(z)$ for a 
COBE-normalized standard-CDM model. The relatively large values for 
this quantity for any source redshift are unrealistic since they are due 
to a mass function of dark halos inconsistent with the cluster abundance.

\begin{figure}
\psfig{file=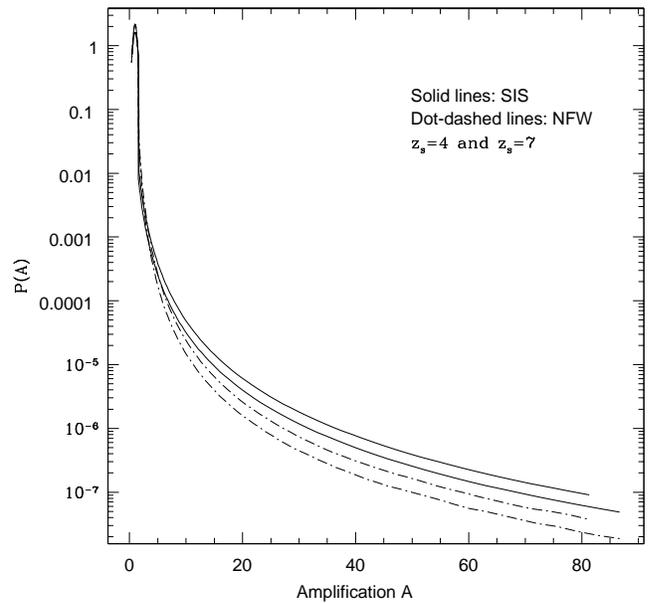,height=4.in}
\caption{Magnification distribution from a population of SIS 
(solid lines) and NFW lenses (dot-dashed lines) for
sources at redshifts $z_{\rm s}=4$ (lower curves) and $z_{\rm s}=7$
(upper curves).}
\label{papicco}
\end{figure}

\begin{figure}
\psfig{file=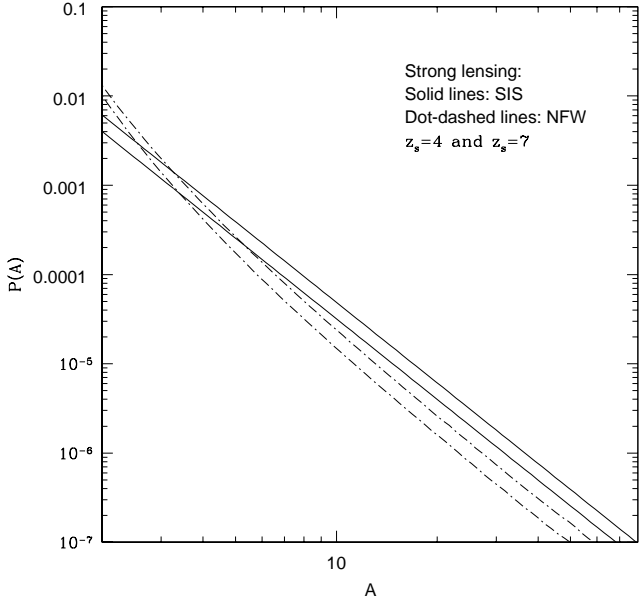,height=4.in}
\caption{High-magnification tail of $P(A)$. Magnifications are plotted
on a logarithmic scale. The plots refer to populations of SIS 
(solid lines) and NFW lenses (dot-dashed lines) for
sources at redshifts $z_{\rm s}=4$ (lower curves) or at $z_{\rm s}=7$
(upper curves).}
\label{pa} 
\end{figure}

\begin{figure}
\psfig{file=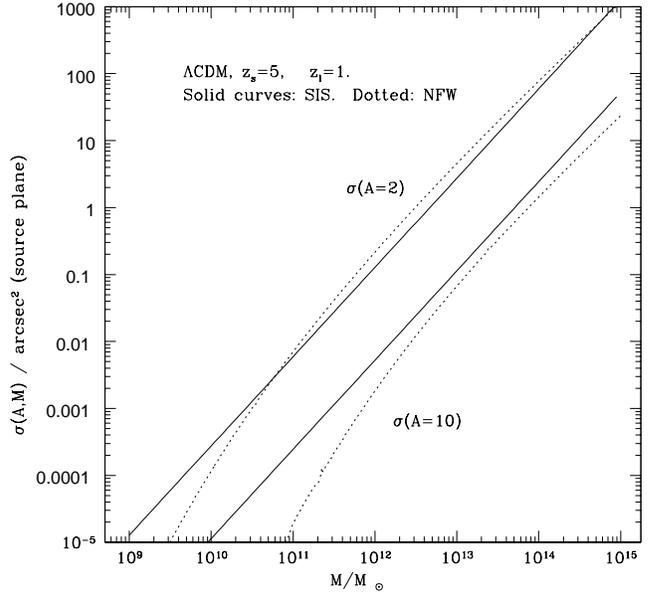,height=4.in}
\caption{Magnification cross sections $\sigma(A)$ (in square
arcsec) for $A>2$ (upper curves) and $A>10$, as a function of the
halo mass. The sources are at $z_{\rm s}=5$ and the lenses at
$z=1$. The cross sections are plotted for SIS (solid lines) and NFW
halos (dotted lines)}.
\label{sigmaA} 
\end{figure}

\begin{figure}
\psfig{file=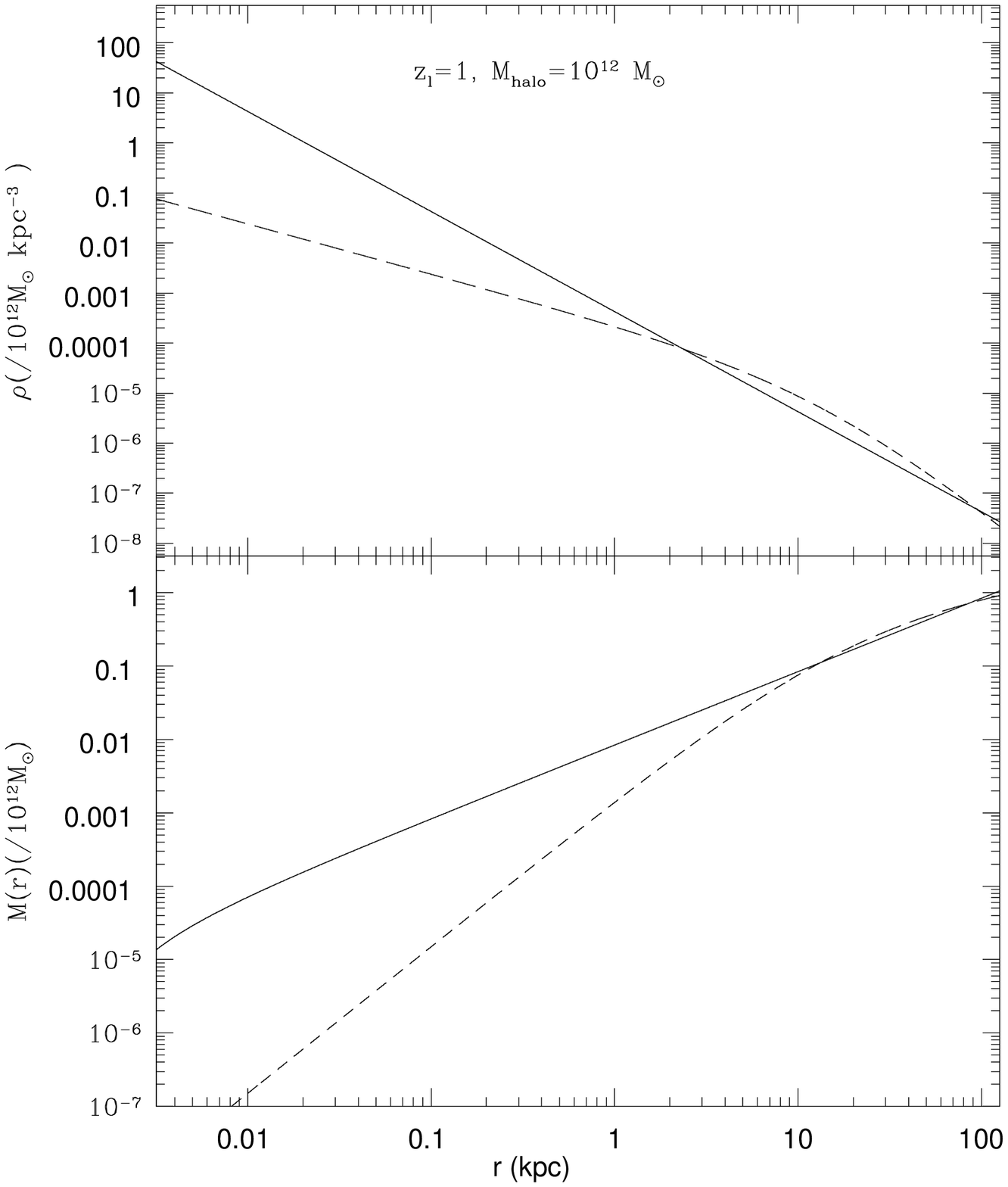,height=4.in}
\caption{Upper panel: density profile for a dark matter halo of
$10^{12}\,h^{-1}\,M_\odot$ at $z=1$. The mass density is plotted
versus the halocentric distance for SIS (solid line) and NFW halos
(dashed line). Lower panel: mass encompassed by radius $r$.}
\label{rhops} 
\end{figure}

\begin{figure}
\psfig{file=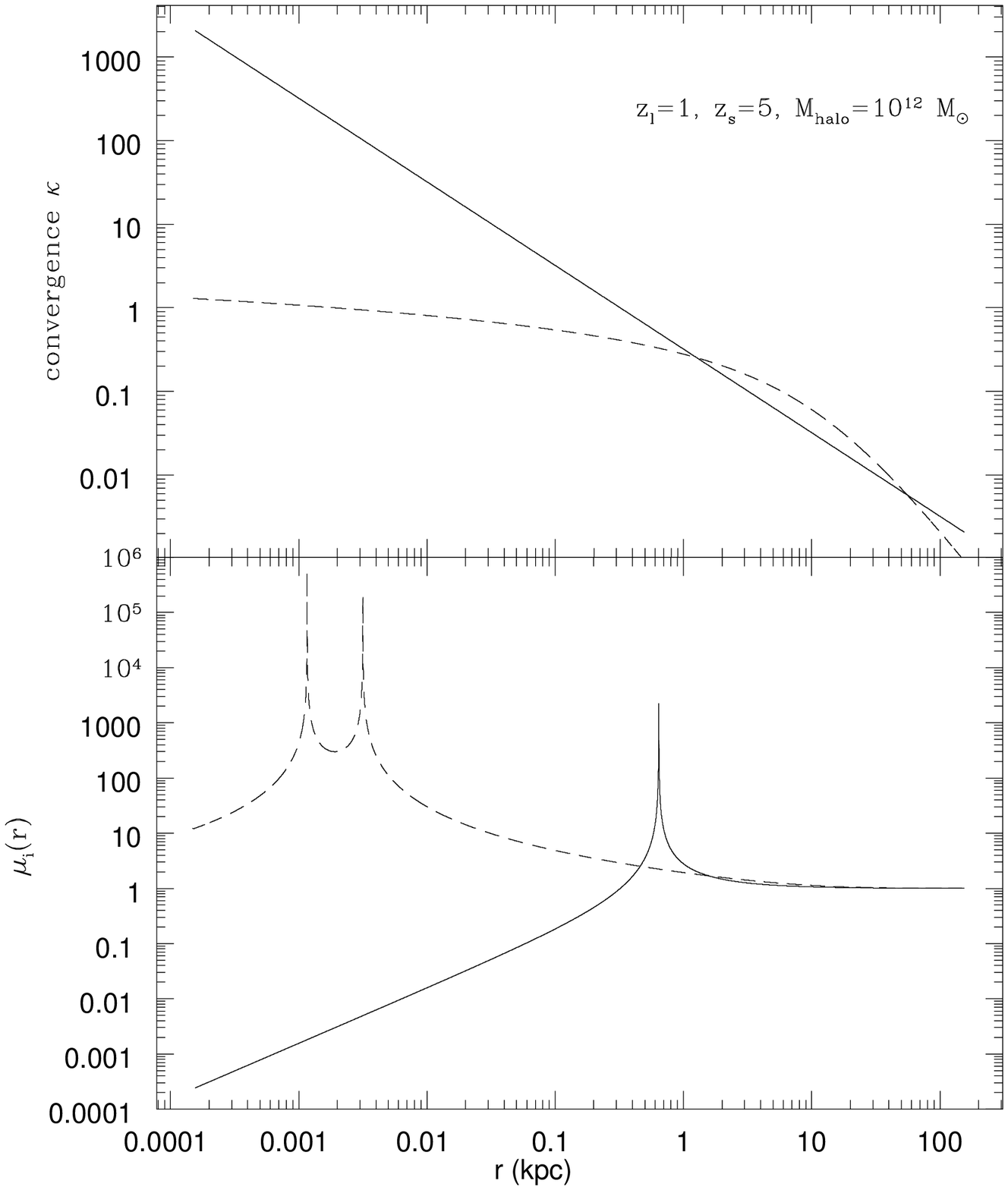,height=4.in}
\caption{Upper panel: convergence $\kappa$ as a function of
halocentric distance for a halo with $M_{\rm
halo}=10^{12}\,h^{-1}\,M_\odot$ at $z_{\rm l}=1$. The sources are at
$z_{\rm s}=5$. The plot refers to a SIS (solid line) or a NFW halo
(dashed line). Lower panel: Magnification as a function of impact
parameter, involving convergence and shear of the gravitational field
(see text).}
\label{arps} 
\end{figure}

\begin{figure}
\psfig{file=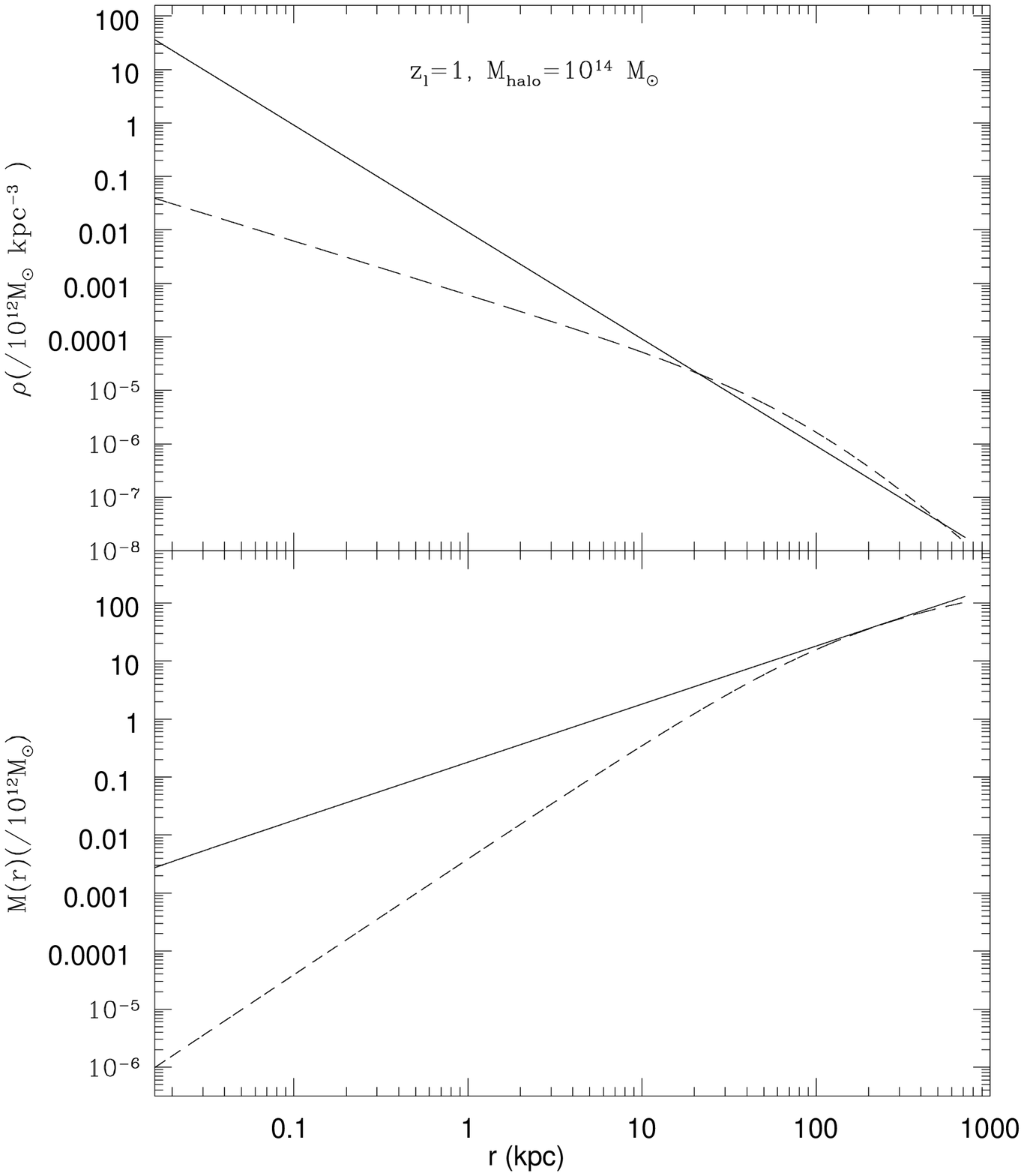,height=4.in}
\caption{Upper panel: density profile for a dark matter halo of
$10^{14}\,h^{-1}\,M_\odot$ at $z=1$. The mass density is plotted
versus the halocentric distance for SIS (solid line) and NFW halos
(dashed line). Lower panel: mass encompassed by radius $r$.}
\label{rho14ps} 
\end{figure}

\begin{figure}
\psfig{file=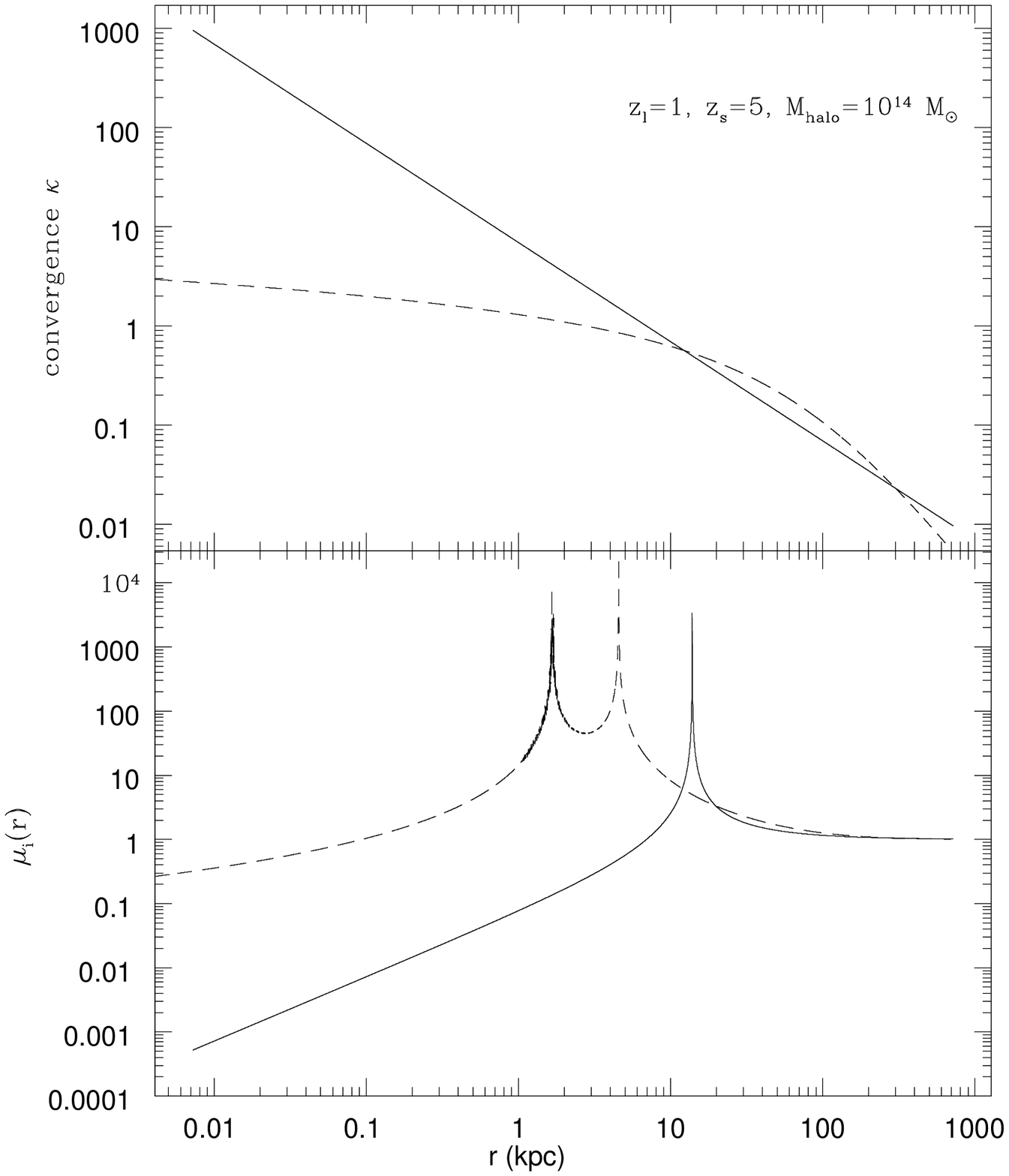,height=4.in}
\caption{Upper panel: convergence $\kappa$ as a function of
halocentric distance for a halo with $M_{\rm
halo}=10^{14}\,h^{-1}\,M_\odot$ at $z_{\rm l}=1$. The sources are
located at $z_{\rm s}=5$. The plot refers to a SIS (solid line) or a
NFW halo (dashed line). Lower panel: magnification as a function of
impact parameter, involving convergence and shear of the gravitational
field.}
\label{ar14ps}
\end{figure}

\subsection{SIS versus NFW profiles}

Figure~\ref{papicco} compares the effect of SIS and NFW lens profiles 
on the magnification distributions for sources at $z_s=4$ and
$z_s=7$, including the weak lensing effect for $A<A_{\rm cut}$.  
The magnification distributions are obtained by integrating 
over lens masses the cross sections 
in the source plane, weighted by the mass distribution [Eq.~(\ref{Sheth})]. The
weak-lensing regime, responsible for magnifications below $A_{\rm
cut}$, gives rise to a Gaussian peak near $A=1$ whose dispersion 
increases with increasing source redshift. In the strong-lensing
regime, the probability distribution is obtained by solving
Eq.~(\ref{lensequation0}) numerically and inserting the result 
in Eq.~(\ref{Prob}). The asymptotic behaviour $\propto
A^{-3}$ is reached earlier by the most distant sources.
The high-magnification tail of Fig.~\ref{papicco} is shown in closer
detail in Fig.~\ref{pa}. Note that the plotted distribution of
magnifications has a discontinuity in $A_{\rm cut }$, i.e. at the 
transition between the weak and strong lensing regimes. 

The two density profiles lead to slightly different magnification
distributions. In particular, the NFW lens is
more efficient than the SIS for moderate magnifications ($2\lsim 
A\lsim 4$), and less efficient for high magnifications. In fact, NFW
lenses have smaller high-magnification cross sections than SIS lenses
of equal mass, even if the average magnification is higher. This can
be read off Fig.~\ref{sigmaA}, where we plot the cross section for
magnifications $A>2$ and $A>10$ as a function of the halo mass for the
two models, keeping fixed the configuration of the system. For
virtually all halo masses, $\sigma(10)_{\rm NFW}<\sigma(10)_{\rm
SIS}$, while $\sigma(2)_{\rm NFW}>\sigma(2)_{\rm SIS}$. Even though
the latter relation fails for very small lens masses, it
still holds when the cross sections are weighted with the appropriate
mass function. As mentioned above, the bulk of the contribution to the 
magnification distribution comes from a limited mass range. 
The effective mass defined
in Sect.~\ref{effect} is nearly equal for SIS and NFW profiles, 
namely $\sim 10^{11-12}M_\odot$. Although
$\langle M\rangle$ depends (albeit weakly) on the lens redshift, massive
clusters never contribute substantially to the integrand in
Eq.~(\ref{Prob}) because they are extremely rare.

The SIS and NFW density profiles and the corresponding mass encompassed 
within the radius $r$ for a $10^{12}\,M_\odot$ object are shown 
in Fig.~\ref{rhops}. Virial radii are approximately equal in the 
two cases. For the NFW profile, the virial and scale radii
are $r_{200}\simeq140\,h^{-1}\,{\rm kpc}$ and $r_{\rm
s}\simeq15\,h^{-1}\,{\rm kpc}$, respectively, for $h=0.65$; 
the concentration is $c\simeq10$. The convergence $\kappa$ for the two
profiles is shown in the upper panel of Fig.~\ref{arps} as a function
of halocentric distance. Unlike the surface density, the convergence
depends also on the geometry of the lens system. Here, the lens is at
$z_{\rm l}=1$ and the source at $z_{\rm s}=5$. The two convergence
profiles are quite similar, but the shear is also playing a
fundamental role to determine the magnification distribution. 
In the lower panel
of Fig.~\ref{arps}, we plot, for the two profiles, $\det^{-1}{\cal A}$, 
i.e.~the image magnification $\mu$ of an image with impact parameter
$r$ in the lens plane. Note that, even if some images are demagnified
by lensing, the total magnification of the source is always $>1$
(e.g.~Schneider et al.~1992). The critical curves behave 
differently. Even though the NFW profile has a singular core, it has
tangential and radial critical curves (Bartelmann 1996), while the SIS
has only a tangential critical curve (whose caustic degenerates to a
point for all axially symmetric lenses). Consequently, the maximum
image number is two for SIS and three for NFW lenses.

Figure~\ref{arps} shows that the tangential critical curve of a SIS occurs at 
a larger radius than both critical curves of an NFW halo with equal mass. This
means that high magnifications are favored in the SIS model, because
the corresponding cross sections in the source plane is larger. On the other 
hand, the SIS profile yields lower total magnifications 
when the source lies well outside the outer caustic in the
source plane. The NFW cross sections for $\mu_{\rm tot}\sim 2$
generally overcome the SIS ones, due to the fact that the
lensing potential is less curved for the NFW than for the SIS profile,
hence this also occurs for more massive halos. In Fig.~\ref{rho14ps}, we
plot the equilibrium configurations for a $10^{14}\,M_\odot$ halo. The
NFW profile has virial and scale radii of
$r_{200}\simeq690\,h^{-1}\,{\rm kpc}$ and $r_{\rm
s}\simeq100\,h^{-1}\,{\rm kpc}$, respectively, for $h=0.65$ (the
concentration is $c\simeq6.5$).

In Fig.~\ref{ar14ps}, we plot the convergence and the magnification
for a single image as a function of its location. After adding all
image magnifications, inverting to find the cross section
$\sigma(\mu)$ vs.~source position, and integrating over all lenses up
to the source redshift, this explains the enhancement of the NFW
probability for low magnifications with respect to the SIS model,
shown in Fig.~\ref{papicco}, and the opposite effect for large
magnifications. However, we see that the probability distributions are
quite similar and the effects of lensing on a source
population turn out to be nearly identical for lenses with the two density
profiles.

We finish with a cautionary note. The simulations resulting in the
density profile of Eq.~(\ref{NFW}) did not have sufficient resolution for
halo masses $\lsim 10^{10}\,h^{-1}\,M_\odot$. We are therefore
extrapolating the validity of this density profile as well
as the relation between halo concentration and mass to a mass range
where higher-resolution $N$-body simulations would be required.
Furthermore, as Porciani \& Madau (2000) pointed out, in order to
obtain agreement with the data on image separations of QSOs one should
consider baryonic cooling in dark matter halos, able to transform NFW
halos into isothermal distributions for masses smaller than some 
threshold. In such a picture, DM halos are modeled as NFW halos only
above a threshold that is certainly well above the smallest mass that
we are considering here.

\subsection{Effects on the source counts}

\begin{figure}
\psfig{file=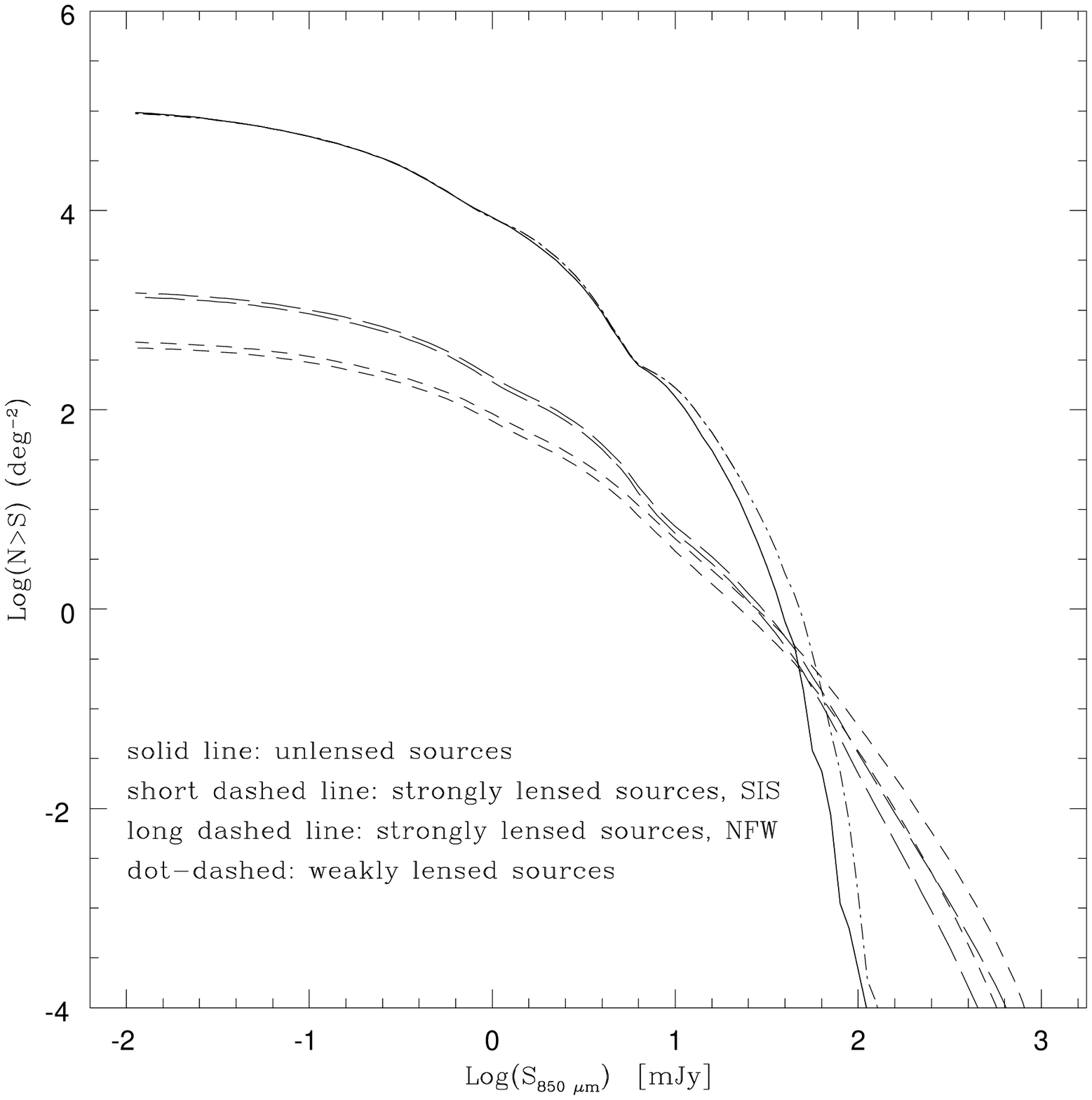,height=4.in}
\caption{Integral Source counts at $850\,\mu$m per square degree.
Unlensed counts are given by the solid line, and counts including weak
lensing by the dot-dashed line.  Short-dashed lines show, for the SIS
model,  strongly-lensed source counts with  $A_{\rm E,max}=  10$ (lower
short-dashed  lines)or $A_{\rm E,max}= 30$ (upper short-dashed lines), as
described in the text. Long-dashed lines are the same for the NFW model
of lenses.}
\label{conteggi}
\end{figure}

We can finally estimate the strongly lensed source counts, and compare
them with the ``unlensed'' and weakly lensed counts, in a given wave
band, based on the model by Granato et al.~(2000). Using the
``corrected'' magnification distribution $p(A)$ for extended sources,
Eq.~(\ref{correction}), we compute lensed counts at $850\,\mu$m, using
SIS and NFW lenses. In Fig.~\ref{conteggi}, the solid line shows the
integral source counts that we expect at $850\,\mu$m from the source
distribution described in Sect.~\ref{galaxy}, ignoring lensing. The
dot-dashed line includes only weak lensing, using the
low-magnification tail of the magnification distribution
$p(A,z)$. Since the latter is modeled as a Gaussian with small
dispersion around the mean $A=1$, we can see from Fig.~\ref{conteggi}
that weak lensing by large-scale structures has very little effect on
the integral source counts, even though the variance of the
distribution increases with source redshift. The weakly lensed counts
are therefore quite similar to the unlensed ones given by Granato et
al.~(2000). Even the weakly lensed counts, however, fall above the
unlensed counts where the number-count function falls most steeply.

The effects of strong lensing from a Sheth \& Tormen mass distribution
of SIS and NFW lenses is plotted in Fig.~\ref{conteggi} as short- and
long-dashed lines, respectively; for both density profiles, 
following the results of section \ref{maxampli}, we show in  
Fig.~\ref{conteggi} the results obtained plugging either $A_{\rm E,max}=
10$ (lower lines)or $A_{\rm E,max}= 30$ (upper lines).

 We can see that the contribution of strong lensing from a SIS model is
here of the order $10^{-2}$ in the flat part of the counts, while it
dramatically raises up in the steepest region, overcoming the weakly
lensed counts. This is due to the very strong magnification bias for
these sources. First, the source redshift, $z\simeq5$, is quite high. The
probability for a source to undergo a lensing event with high
magnification increases with increasing source redshift. Second, and
importantly, the steep source counts, discussed in Sect.~\ref{galaxy} and
by Blain (1996), provide a huge reservoir of source to be magnified above
the flux limit of the observation.

At flux densities around $100\,$mJy, the counts will be dominated by
sources magnified by a factor $A\sim10-20$.
As we have seen, such quite high magnifications are less probable for the
NFW lenses than for the SIS population, which explains the relative
values of strongly lensed source counts from NFW and SIS lenses at flux
densities  $\gsim 100\,$mJy of Fig.~\ref{conteggi}. 
At higher flux densities, lensed sources  are depressed by the effect of
having taken a maximum amplification $A_{\rm E,max}= 30$.    

The substantial magnification bias at  flux densities around
$100\,$mJy allows to detect sources otherwise not bright enough for 
detection.  This result may have very  interesting consequences on the 
expected number of SCUBA sources to be found in the whole sky maps 
produced by the incoming  Planck Surveyor Satellite, operating  at nine 
frequency channels between 0.3 and 10 mm (Mandolesi et al. 1998).
The detailed study of this issue will be the subject of a forthcoming
work (Perrotta et al.~2001).

\section*{Acknowledgements}

We are grateful to L. Danese, L. Moscardini, S. Matarrese, C. Porciani, 
A. Blain and C. Lacey for useful suggestions and discussions. 
We remember with gratitude R. De Ritis, and, together with
him, we acknowledge E. Piedipalumbo, M. Demianski and A.A. Marino,
who kindly provided the coefficients for the approximation of the
solution to the Dyer-Roeder equation used in this paper.
F.P. wishes to thank the MPA  for
kind hospitality.

Work supported in part by MURST and ASI.

\end{document}